\documentclass[aps,pre,reprint,noshowpacs,twocolumn,groupedaddress]{revtex4}

\usepackage{hyperref}
\usepackage{amsmath}
\usepackage{amsfonts}
\usepackage{amssymb}
\usepackage{graphicx}
\usepackage{empheq}
\usepackage{diagbox}

\usepackage{tikz,xcolor,hyperref}

\definecolor{linkcolour}{HTML}{000066}	
\hypersetup{colorlinks,breaklinks,
	urlcolor=linkcolour, 
	linkcolor=linkcolour,
	citecolor=linkcolour}

\DeclareMathOperator*{\Res}{Res}

\definecolor{lime}{HTML}{A6CE39}
\DeclareRobustCommand{\orcidicon}{
	\begin{tikzpicture}
		\draw[lime, fill=lime] (0,0) 
		circle [radius=0.16] 
		node[white] {{\fontfamily{qag}\selectfont \tiny ID}};
		\draw[white, fill=white] (-0.0625,0.095) 
		circle [radius=0.007];
	\end{tikzpicture}
	\hspace{-2mm}
}

\newcommand{\orcidauthorNB}{\href{https://orcid.org/0000-0002-1554-3820}{\orcidicon}} 
\newcommand{\orcidauthorMRF}{\href{https://orcid.org/0000-0001-5864-9636}{\orcidicon}} 
\newcommand{\orcidauthorKY}{\href{https://orcid.org/0009-0001-1171-3052}{\orcidicon}} 

\hypersetup{colorlinks=true, urlcolor=blue, linkcolor=blue, citecolor=blue,plainpages=false}

\begin{document}
	\title{Generalized Wigner–Smith theory for perturbations at exceptional and diabolic point degeneracies}
	
	\author{Kaiyuan Wang\orcidauthorKY$^{1,2}$, Niall Byrnes\orcidauthorNB$^1$ and Matthew R. Foreman\orcidauthorMRF$^{1,2}$}
		\email[]{matthew.foreman@ntu.edu.sg}
	\affiliation{
		$^1$School of Electrical and Electronic Engineering, Nanyang Technological University, 50 Nanyang Avenue, Singapore 639798 \\
		$^2$Institute for Digital Molecular Analytics and Science, 59 Nanyang Drive, Singapore 636921
	}
	\date{\today}
	
\begin{abstract}
Resonant spectral degeneracies, including diabolic (DP) and exceptional (EP) points, exhibit unique sensitivity to external perturbations, enabling powerful control and engineering of wave phenomena. We present a residue-based perturbation theory that quantifies complex resonance splitting of DP and EP type spectral degeneracies using generalized Wigner–Smith operators. We validate our theory using both analytic Hamiltonian models and numerical electromagnetic simulations, demonstrating excellent agreement across a range of cases. Our approach accurately predicts degenerate resonance splitting using only scattering data, offering a powerful framework for sensitivity analysis, inverse design, and quantitative sensing.
\end{abstract}
	
\maketitle
The Wigner–Smith (WS) time delay operator, initially introduced in quantum scattering theory \cite{Wigner1955Lower, Smith1960Lifetime}, serves as a fundamental tool for characterizing the temporal response of resonant systems. For a system described by a scattering matrix $\mathbf{S}(\omega)$, the real part of the eigenvalues of the WS matrix $\mathbf{Q}_\omega= -i\, \mathbf{S}^{-1}\ \partial_\omega \mathbf{S}$ quantify the group delays of a set of incident wavepackets with modal structures given by the corresponding eigenvectors \cite{9142355}. $\mathbf{Q}_\omega$ connects a system's frequency-domain response to its energetic and temporal properties, allowing studies of, e.g., time-delay statistics in chaotic cavities \cite{Grabsch2018}, light storage in random media \cite{Durand2019Optimizing}, and density of states enhancement in nuclear physics \cite{Higgins2020Nonresonant}. More recently, generalized WS (GWS) operators of the form $\mathbf{Q}_\xi = -i\, \mathbf{S}^{-1} \partial_\xi \mathbf{S}$, where $\xi$ denotes an arbitrary system parameter, have been proposed, extending conventional WS operators into parameter space and connecting external scattering observables to internal parametric sensitivities \cite{horodynski2020optimal}. A key feature of GWS operators is that their principal modes, defined in analogy to the dispersion-free principal modes derived from $\mathbf{Q}_\omega$ \cite{Carpenter2015Observation}, have been shown to exhibit insensitivity to the variable conjugate to $\xi$ \cite{Ambichl2017}. This insight has led to a wealth of  advances in wavefront shaping and control \cite{DelHougne2021Coherent}, including targeted focusing in complex media \cite{sol2025blind}, guided light delivery in deformed optical fibers \cite{matthes2021learning}, manipulation of optical forces and trap stiffness in microparticle systems \cite{butaite2024photon}, tailored energy redistribution in time-varying systems \cite{huepfl2023cooling}, and the construction of optimal information states for scattering measurements \cite{bouchet2021maximum}.

Recent studies have extended GWS operators to describe parametric spectral shifts in non-Hermitian systems \cite{Byrnes2024a, Byrnes2024b}. In particular, Byrnes and Foreman~\cite{Byrnes2024a} demonstrated that a complex analytic treatment of the GWS operator yields compact perturbative formulae for shifts in isolated resonances or anti-resonances associated with the poles and zeros of $\mathbf{S}$. Their treatment, however, was restricted to non-degenerate  resonances associated with \emph{simple} poles of $\mathbf{S}$ and therefore fails at diabolic points (DPs), where multiple resonant modes share the same frequency \cite{berry1984diabolical},  and exceptional points (EPs), where  resonant modes degenerate in frequency also become linearly dependent \cite{Miri2019}. These degeneracies are not only mathematically rich, but also practically relevant. For example, DPs underpin mode-splitting sensors \cite{zhu2010onchip}, while EPs enable chiral mode control, topological transport, and enhanced sensitivity \cite{Lai2019, Li2023StochasticEP, PhysRevLett.132.243601}. A GWS framework that rigorously handles degenerate resonances remains lacking. In this work, we address this gap by developing a residue-based formalism that extends earlier perturbative formulae to arbitrary-order DP and EP type degeneracies and naturally reduces to earlier results in the non-degenerate case. We validate our derived formula using both analytic Hamiltonian models and electromagnetic simulations, thereby demonstrating its applicability to  nanophotonic structures.

\emph{GWS perturbation theory at degenerate poles--} We begin by briefly reviewing the key concepts from non-Hermitian physics most relevant to our study. More information can be found in the supplemental material and the literature \cite{Ashida2020}. Consider a physical system with a non-Hermitian $N\times N$ Hamiltonian matrix $\mathbf{H}(\alpha)\in\mathbb{C}^{N\times N}$ depending on a complex control parameter $\alpha$. Our work concerns \emph{resonant} DP and EPs associated with the eigenmodes of $\mathbf{H}$, as opposed to \emph{scattering} DP and EPs associated with those of $\mathbf{S}$ \cite{Sweeney2019}. We assume that for a particular parameter value $\alpha = \alpha_0$, $\mathbf{H}(\alpha_0)$ has a repeated (degenerate) eigenvalue $\omega_p$ with algebraic multiplicity $\mathrm{AM}=N$ and geometric multiplicity $\mathrm{GM}$ satisfying $1 \leq \mathrm{GM} \leq N$. When $\mathrm{AM} = \mathrm{GM}$, $\mathbf{H}(\alpha_0)$ is diagonalizable and the degeneracy corresponds to a DP. When $\mathrm{GM} < \mathrm{AM}$, however, $\mathbf{H}(\alpha_0)$ is defective and contains at least one Jordan block of size greater than one. For our purposes, an EP will correspond to the case $\mathrm{GM} =1$, meaning $\mathbf{H}(\alpha_0)$ contains a single Jordan block of size $N \times N$. Intermediate cases $1 < \mathrm{GM} < N$ correspond to hybrid DP-EP scenarios as discussed below.

Suppose first that the system possesses an EP and is perturbed so that $\alpha$ shifts from $\alpha_0$ to a nearby value. We assume that this perturbation is generic, meaning the degeneracy is lifted and $\omega_p$ splits into $N$ distinct eigenvalues $\omega_n$ ($n=1,\dots,N$). At a point $\alpha$ in the neighborhood of $\alpha_0$, the perturbed eigenfrequencies admit a Puiseux expansion \cite{kato2013perturbation}
\begin{equation}
	\label{eq:Puiseux}
	\omega_n(\alpha)=\omega_p+c\,e^{2\pi i n/N}\Delta\alpha^{1/N}
	+\mathcal{O}\bigl(\Delta\alpha^{2/N}\bigr),
\end{equation}
where $\Delta\alpha=\alpha-\alpha_0$ and $c\in\mathbb{C}\neq0$. We note immediately that Eq.~(\ref{eq:Puiseux}) implies that, to leading order, the degenerate eigenvalue splits symmetrically and shifts in proportion to $\Delta \alpha^{1/N}$. Deviations from this scaling have been reported for systems experiencing non-generic perturbations, such as those preserving symmetry \cite{Grom2025}. Such cases are beyond the scope of this work.

In addition to $\Delta \alpha$, the shift sensitivity is also governed by the coefficient $c$, which remains to be determined. It has been shown elsewhere that this coefficient is related to the orthogonality of the eigenvectors of $\mathbf{H}(\alpha_0)$ \cite{PhysRevResearch.5.033042}. Here we show that it can also be related to the scattering matrix $\mathbf{S} \in \mathbb{C}^{M\times M}$. A commonly used expression for $\mathbf{S}$ in terms of $\mathbf{H}$ is given by
\cite{10.1063/1.531919}
\begin{equation}
	\mathbf{S}(\omega,\alpha)=\mathbf{I}_{M}
	-i\,\mathbf{W}^{\dagger}
	\bigl(\omega\mathbf{I}_{N}-\mathbf{H}(\alpha)\bigr)^{-1}
	\mathbf{W},\label{eq:mahauxS}
\end{equation}
where $\mathbf{I}_M$ is the $M\times M$ ($N \times N$) identity matrix and $\mathbf{W} \in \mathbb{C}^{N \times M}$ couples the $N$ internal modes of the system to $M$ external scattering channels. A key property of Eq.~(\ref{eq:mahauxS}) is that  degenerate eigenfrequencies of $\mathbf{H}$ are mapped to degenerate poles of $\mathbf{S}$ via the resolvent $(\omega\mathbf{I}_N - \mathbf{H}(\alpha))^{-1}$. It has also been shown that \cite{welters2011jordan}
\begin{equation}
	\label{eq:cNdet}
	c^{N}=-\left.
	\frac{\partial}{\partial\alpha}\,
	\det\bigl[\omega\mathbf{I}_{N}-\mathbf{H}(\alpha)\bigr]
	\right|_{{\omega = \omega_p,~\alpha = \alpha_0}},
\end{equation}
which can be linked to Eq.~(\ref{eq:mahauxS}) using the identity
\begin{equation}
	\det\bigl[\omega\mathbf{I}_{N}-\mathbf{H}(\alpha)\bigr]=\det\bigl[\mathbf{S}^{-1}(\omega,\alpha)\bigr] g(\omega, 
	\alpha), \label{eq:det1}
\end{equation}
where $g(\omega, \alpha)= \det\bigl[\omega\mathbf{I}_{N}-\mathbf{H}(\alpha)-i\mathbf{W}\mathbf{W}^{\dagger}\bigr]$. A short derivation of Eq.~(\ref{eq:det1}) is given in the supplemental material.
Differentiating \eqref{eq:det1} with respect to $\alpha$ and applying Jacobi’s identity to the derivative of $\det (\mathbf{S}^{-1})$, we can obtain $c^N$ by evaluating the limit
\begin{equation}
	\label{eq:cNtrace}
	c^{N}= \left.\lim_{\omega\to\omega_p}
	(\omega-\omega_p)^{N}\bigg(\,i\,\mathrm{tr}\bigl[\mathbf{Q}_{\alpha}(\omega,\alpha)\bigr] -  \frac{1}{g}\frac{\partial g}{\partial \alpha}\bigg)\right|_{\substack{\alpha = \alpha_0}}, 
\end{equation}
where we have used $\det[\omega\mathbf{I}_{N}-\mathbf{H}(\alpha_0)] = (
\omega - 
\omega_p)^N$. To proceed, we note that the final term in Eq.~(\ref{eq:cNtrace}) vanishes provided that $\mathbf{W}$ is non-zero (see supplemental material). Standard limit laws allow us to write 
$
c=\mathrm{Res}_{\omega=\omega_p}
\{[i\,\mathrm{tr}\,\mathbf{Q}_{\alpha}(\omega,\alpha_0)]^{1/N}\}$, 
which can be substituted back into \eqref{eq:Puiseux}. Neglecting higher order terms  yields the resonance shift formula
\begin{equation}
	\label{eq:domegaFinal}
	\Delta\omega_n=e^{2\pi i n/N}\;
	\Delta\alpha^{1/N}\;
	\mathrm{Res}_{\omega=\omega_p}
	\Bigl\{\bigl[i\,\mathrm{tr}\,\mathbf{Q}_{\alpha}(\omega, \alpha_0)\bigr]^{1/N}
	\Bigr\},
\end{equation}
where $\Delta \omega_n = \omega_n - \omega_p$. Notably, \eqref{eq:domegaFinal} exhibits the correct fractional power scaling and reduces to the non-degenerate pole shift formula when $N=1$ \cite{Byrnes2024a,Byrnes2024b}. It should also be noted that it is easy to show, e.g. by considering $\mathbf{S}^{-1}$ instead of $\mathbf{S}$, that the same analysis equally applies to the splitting of  degenerate scattering zeros, such as those corresponding to coherent perfect absorption EPs \cite{wang2021coherent}.

We now turn our attention to DPs, where $\mathrm{GM} = N$ and the Jordan normal form of $\mathbf{H}(\alpha_0)$ is diagonal with repeated entry $\omega_p$. In contrast to EPs, where eigenvalue shifts are strongly correlated, the linear independence of modes at a DP allows the eigenvalues to shift independently, each admitting their own Taylor expansion. Consequently, \eqref{eq:domegaFinal} can not recover all frequency shifts and an alternative approach is required. In principle, the $N$ frequency shifts could be determined by constructing GWS operators from a suitable set of $N$ scattering functions, each isolating a single pole associated with one of the DP modes. Application of the non-degenerate theory to each of these operators would then yield the $N$ desired frequency shifts. The feasibility of this approach, however, depends intricately on the form of $\mathbf{W}$ and $\mathbf{H}$. The aforementioned full rank condition for $\mathbf{W}$, particularly when $M \geq N$ (at least as many scattering channels as internal modes), ensures that $\mathbf{S}$ possesses $N$ non-zero eigenvalues. These eigenvalues, which can easily be extracted from $\mathbf{S}$, are natural candidates for the desired set of scattering functions. In general, however, a single eigenvalue of $\mathbf{S}$ can depend on multiple eigenfrequencies of $\mathbf{H}$ and additional constraints are required to ensure that the poles are cleanly separated.

One simple class of $\mathbf{W}$ matrices for which the eigenvalues of $\mathbf{H}$ do not mix among those of $\mathbf{S}$ can be found by observing the algebraic structure of Eq.~(\ref{eq:mahauxS}). Note first that although $\mathbf{W}$ is not in general square, it possess a right inverse $\mathbf{W}_R = \mathbf{W}^\dagger(\mathbf{W}\mathbf{W}^\dagger)^{-1}$. If $\mathbf{W}\mathbf{W}^\dagger = \gamma\mathbf{I}_N$ for some constant $\gamma$, then it follows that $\mathbf{W}^\dagger = \gamma\mathbf{W}_R$ and Eq.~(\ref{eq:mahauxS}) thus expresses pseudo-similarity between $\mathbf{S}$ and $\mathbf{H}$. Physically, this condition dictates that each internal mode couples to the exterior through orthogonal, non-interfering channels with uniform coupling strength $\gamma$. The eigenvalues of $\mathbf{S}$ will then be of the form $\mu_n = 1 - i\gamma(\omega - \omega_n)^{-1}$ and analysis of the operators $Q_{\alpha,n} = -i\mu_n^{-1}\partial_\alpha\mu_n$ yields the pole shifts. Though this condition is sufficient to avoid mixing, it is not necessary, and pole mixing can be avoided even in cases where $\mathbf{W}\mathbf{W}^\dagger$ is not diagonal. An example of such a case is presented in the supplemental material, but an exhaustive analysis is beyond the scope of this paper. In any case, we note that if the form of $\mathbf{W}$ is known, one can always compute its inverse directly and obtain the resolvent $(\omega\mathbf{I}_{N}-\mathbf{H})^{-1} = -i\mathbf{W}_R^{\dagger}(\mathbf{S} - \mathbf{I}_M)\mathbf{W}_R$,
whose eigenvalues each are of the form $(\omega - \omega_n)^{-1}$. This calculation reverts the problematic mixing caused by $\mathbf{W}$, yielding a set of functions that can be used to find the pole shifts. As a final remark, we note that the cases $1 < \mathrm{GM} < N$ present similar challenges associated with the structure of 
$\mathbf{W}$. In principle, however, the analysis can be restricted to each 
Jordan block individually, allowing the techniques described to be applied to each one.

\emph{EP and DP in a two-level system--}To verify our theory, we now apply it to the analysis of perturbations to a canonical, non-Hermitian model widely used to describe coupled microring resonators and similar photonic dimers~\cite{Hodaei2017}.  This system is characterized by the $2\times2$ effective Hamiltonian 
\begin{align}
	\mathbf H(\kappa)=
	\begin{pmatrix}
		\omega_1-i\gamma_1/2 & \kappa\\
		\kappa & \omega_2-i\gamma_2/2
	\end{pmatrix}, \label{eq:Heff}
\end{align}
where $\omega_1$ and $\omega_2$ are the resonant frequencies of the two modes, $\gamma_1$ and $\gamma_2$ represent their loss rates, and $\kappa$ is the complex coupling amplitude. For convenience, the detuning and loss contrast are defined as $\Omega\equiv\omega_1-\omega_2$ and $\Gamma\equiv\gamma_1-\gamma_2$ respectively. An EP  arises when $\kappa = \kappa_\mathrm{EP}$ is such that $\det[\omega \mathbf{I} - \mathbf{H}(\kappa_\mathrm{EP})
] = 0$ has a repeated root $\omega = \omega_{\mathrm{EP}}$, which occurs when $\kappa_{\mathrm{EP}}
= \pm i(\Omega/2-i\Gamma/4)$ and $\omega_\mathrm{EP}
= (\omega_1+\omega_2)/2-i(\gamma_1+\gamma_2)/4$. For a small deviation  
$\kappa=\kappa_{\mathrm{EP}}+\delta\kappa$ with $|\delta\kappa|\!\ll\!|\kappa_{\mathrm{EP}}|$, direct diagonalization of the resulting $\mathbf H$ yields the leading-order eigen-splitting
\begin{equation}
	\Delta\omega_{\pm}\simeq
	\pm\sqrt{2\,\kappa_{\mathrm{EP}}\,\delta\kappa} .
	\label{eq:dimer_diag_split}
\end{equation}
To compare \eqref{eq:dimer_diag_split} to \eqref{eq:domegaFinal} for the dimer system, we must first define the form of the channel matrix $\mathbf{W}$. For simplicity, we choose  
$\mathbf W=\mathrm{diag}\!\bigl(\sqrt{\gamma_{1c}},\sqrt{\gamma_{2c}}\bigr)$,
which couples each internal mode to an independent external channel with rates $\gamma_{1c}$ and $\gamma_{2c}$. The introduction of $\mathbf{W}$ requires us to modify the effective Hamiltonian according to $\mathbf{H} \to \mathbf{H} -i\mathbf{W}\mathbf{W}^\dagger/2$, which, given the form of $\mathbf{W}$, is equivalent to redefining the loss rates in Eq.~(\ref{eq:Heff}) according to $\gamma_i \to \gamma_i + \gamma_{ic}$. With these redefinitions, the forms of $\kappa_\mathrm{EP}$ and $\omega_{\mathrm{EP}}$ as given above remain invariant. Using Eq.~(\ref{eq:mahauxS}), the scattering matrix and corresponding GWS operator $\mathbf{Q}_\kappa$ can be computed. After some algebra, we find
\begin{align}
	\Res_{\omega=\omega_0}\left[\sqrt{i\,\mathrm{tr}\,\mathbf Q_\kappa(\kappa_\mathrm{EP})}\right]       =\sqrt{2\kappa_{\rm EP}},
\end{align}
which, when substituted into ~\eqref{eq:domegaFinal} with $N=2$, yields $
\Delta\omega_{\pm}
= \pm\sqrt{2\,\kappa_{\mathrm{EP}}\,\delta\kappa}
$, in exact agreement with  \eqref{eq:dimer_diag_split}.

By adjusting the parameters in Eq.~(\ref{eq:Heff}), we can construct a DP within the same mathematical framework. This can be achieved by setting $\Omega=\Gamma=0$, which reduces the effective Hamiltonain to $\mathbf{H} = (\omega_0 -i\gamma/2)\mathbf{I}_2 + \kappa\mathbf{J}_2$, where $\mathbf{J}_2$ is the $2\times 2$ exchange matrix and $\omega_0 =\omega_1 (= \omega_2)$. This case now trivially admits a DP at $\kappa_\mathrm{DP} = 0$ with repeated eigenvalue $\omega_0 -i\gamma/2$. Altering $\kappa$ by $\delta\kappa$ shifts the eigenvalues linearly by $\Delta\omega_{\pm} = \mp\delta\kappa$. At the same time, with $\mathbf{W}$ defined as before, but with $\gamma_{1c}=\gamma_{2c}=\gamma_c$ and $\gamma \to \gamma + \gamma_{c}$, it is straightforward to show that the eigenvalues of $\mathbf{S}$ are given by $\mu_\pm = 1 -i\gamma_c(\omega - \omega_0 + i\gamma/2\pm\kappa)^{-1}$. Note that this form of $\mathbf{W}$ satisfies the property $\mathbf{W}\mathbf{W}^\dagger = \gamma\mathbf{I}_2$ as discussed above and does not result in pole mixing among $\mu_\pm$. The GWS operator $Q_{\kappa,\pm}=-i\mu_{\pm}^{-1}\partial_\kappa\mu_{\pm}$ is then given by
\begin{align}
	Q_{\kappa,\pm}= \frac{\pm\gamma_c}{(\omega-\omega_0 +i\gamma/2 \pm \kappa)(\omega-\omega_0 +i\gamma/2 \pm \kappa -i\gamma_c)},
\end{align}
which, after partial fraction decomposition, can be shown to have residue $\pm i$ at $\omega = \omega_0 -i\gamma/2$ when $\kappa=0$. Eq.~(\ref{eq:domegaFinal}) with $N=1$ therefore predicts $\Delta\omega_{\pm} = \mp \delta\kappa$ as expected.

\begin{figure}[t]
	\centering
	\includegraphics[width=\linewidth]{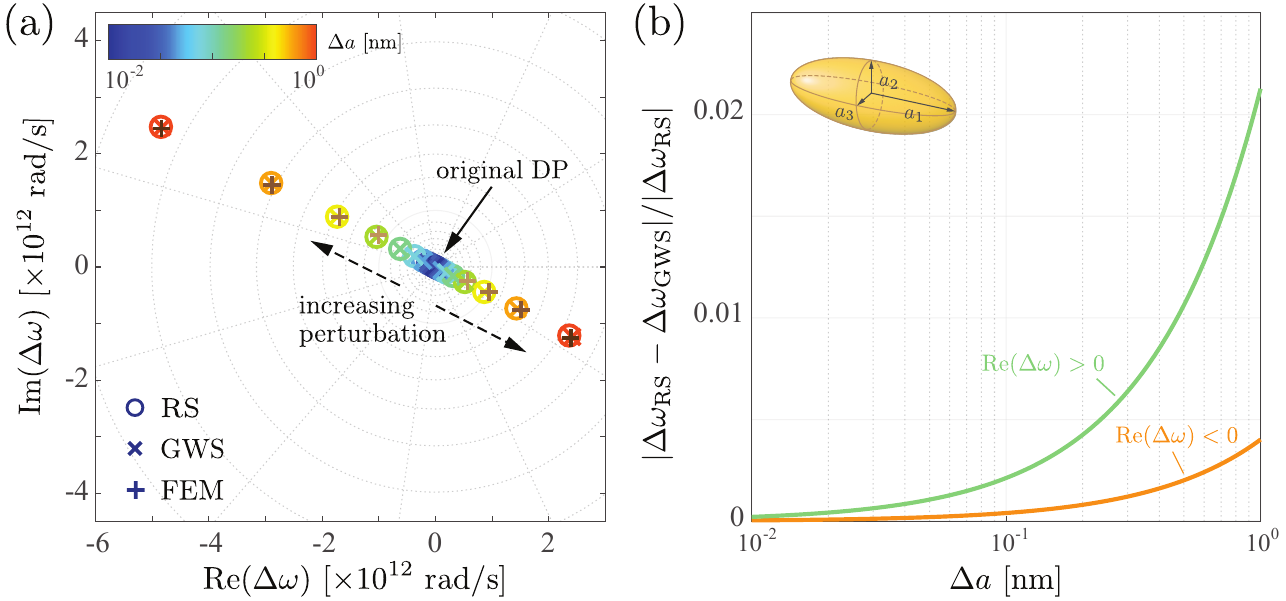}
	\caption{\textbf{(a)} Trajectories of poles in the polarizability of an initially spherical nanoparticle subjected to an axial elongation (inset) along its first principal axis ($a_1 \rightarrow a_1 + \Delta a$) as computed by root searching (RS, $\circ$ markers), our GWS residue formula ($\times$ markers), and full FEM simulations ($+$ markers).  Colors indicate perturbation amplitude $\Delta a$ on a logarithmic scale. \textbf{(b)} Magnitude of the relative difference in pole shifts for both branches as obtained from the RS and GWS methods as a function of $\Delta a$.}
	\label{fig:ellipsoid_shifts}
\end{figure}

\emph{DP in ellipsoidal nanoparticles--}We now consider the case of a DP in a small, homogeneous, metallic ellipsoid. Our ellipsoid, embedded in a background of unity permittivity, has semi-axes $a_1,a_2,a_3$ (see inset of Figure~\ref{fig:ellipsoid_shifts}) and complex permittivity $\varepsilon(\omega)$ taken as that of gold assuming a Drude–Lorentz model \cite{rakic1998optical}. In a coordinate system aligned with the ellipsoid's principal axes, the $3\times 3$ polarizability tensor is diagonal with entries  \cite{bohren2008absorption} 
\begin{align}
	\mathcal{P}_i(\omega)\sim\frac{\varepsilon(\omega)-1}{1+L_i[\varepsilon(\omega)-1]},
\end{align}
where $L_i$ are geometrical factors. In the case of a sphere ($a_1=a_2=a_3$), $L_1=L_2=L_3=1/3$ and the principal polarizabilities become degenerate, each exhibiting a pole $\omega_p$ at the Fr\"ohlich resonance condition $\varepsilon(\omega_p) =-2$. The sphere therefore constitutes a DP in the space of ellipsoid shapes with the polarizability tensor playing the role of the scattering matrix.

A small axial perturbation to the $j$'th semi-axis of the sphere, $a_j\rightarrow a_j+\Delta a_j$, breaks the geometric symmetry, partially lifting the degeneracy. The pole shifts in each polarizibility component can be found by analyzing the functions $Q_{\Delta a_j, i} = -i\mathcal{P}_i^{-1}\partial_{\Delta a_j}\mathcal{P}_i$, from which we find that the pole in $\mathcal{P}_i$, $\omega_{p,i}$, shifts according to
\begin{align}
	\Delta\omega_{p,i} =\Delta a_j\frac{6(1-3\delta_{ij})}{5a}\Res_{\omega=\omega_p}\bigg(\frac{1}{\varepsilon(\omega)+2}\bigg),\label{eq:ellipsshift}
\end{align}
where $a$ is the sphere radius. Notably, Eq.~(\ref{eq:ellipsshift}) shows that the pole associated with the elongated axis shifts in the opposite direction to those of the other two axes, with twice the magnitude. The polarizabilities of the remaining axes stay degenerate, as their symmetry is preserved. To test the validity of Eq.~(\ref{eq:ellipsshift}), we compared it to the results of two additional methods: root-tracking of the inverse of the quasistatic polarizability, and numerical tracking of the poles using finite-element (FEM) simulations in COMSOL. In particular, we considered shifts in a degenerate localized surface plasmon resonance at $\omega_p \simeq (3700 - 400i) \times 10^{12}\,\,\mathrm{rad/s}$ in a sphere of radius $50$ nm, stretched along one axis with $\Delta a$ ranging from $10^{-2}$ to $1$ nm in 10 log-spaced intervals. Figure~\ref{fig:ellipsoid_shifts}(a) compares the computed complex frequency shifts. As can be seen, all three methods show good quantitative agreement across the range of perturbations considered. The pole splitting is clearly visible, with the poles traversing two oppositely directed branches. The upper-left branch ($\mathrm{Re}(\Delta\omega) < 0$) corresponds to $i=j$ (the pole in the polarizability of the perturbed axis) and features points spaced twice as far apart as those on the doubly degenerate, lower-right branch ($\mathrm{Re}(\Delta\omega) > 0$), which corresponds to $i\neq j$ (poles in polarizabilities of the unperturbed axes). Figure \ref{fig:ellipsoid_shifts}(b) shows the relative numerical differences between the GWS and numerical root based approaches, confirming the agreement seen in (a). Errors for both branches were found to grow approximately quadratically, indicating that the dominant source was the neglect of higher-order terms in the perturbative expansion.

\begin{figure}[t]
	\centering
	\includegraphics[width=\linewidth]{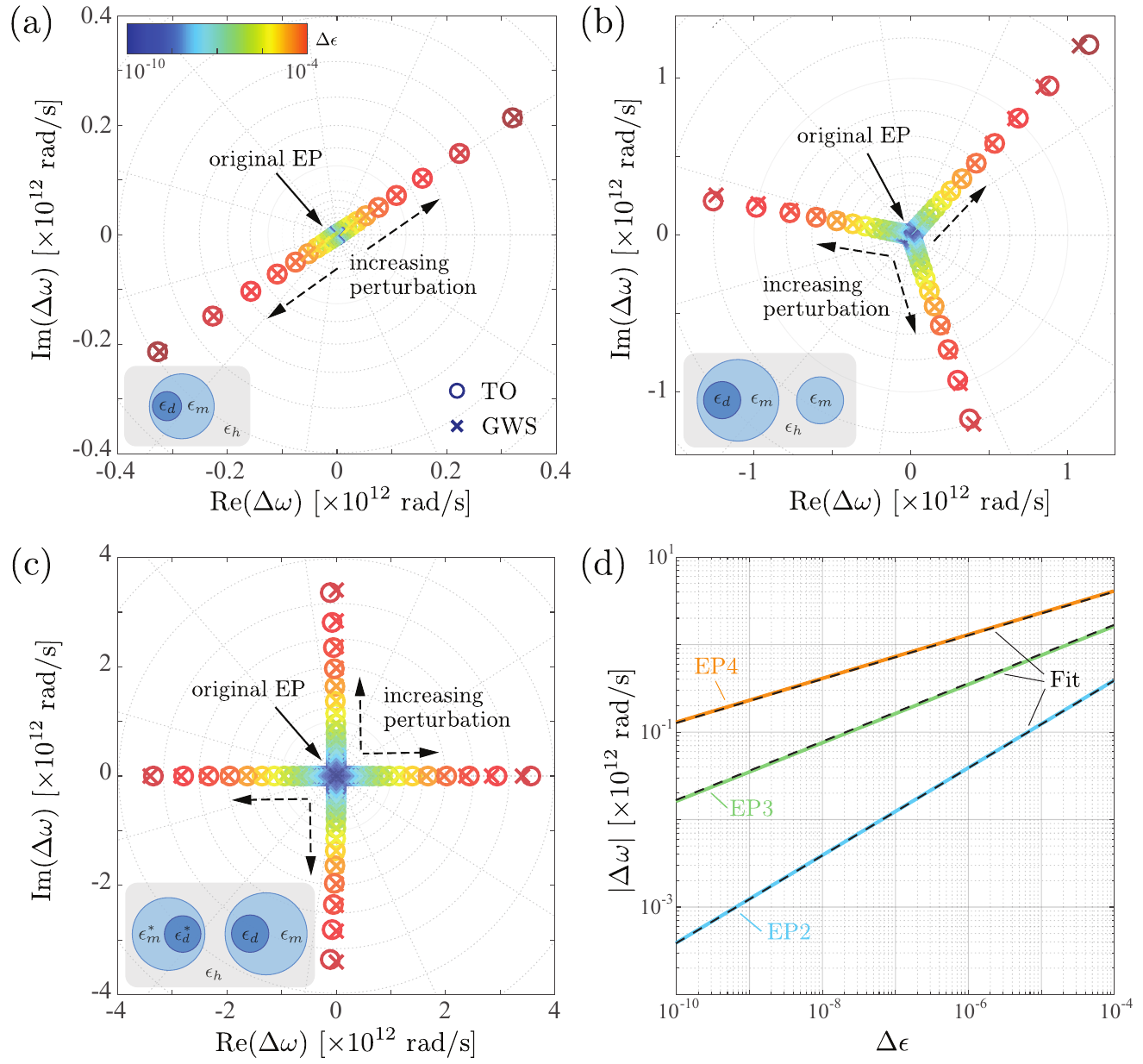}
	\caption{Splitting in the complex plane of \textbf{(a)} EP$_2$, \textbf{(b)} EP$_3$, and \textbf{(c)} EP$_4$ degenerate resonances supported in TO designed plasmonic nanowire structures (insets) upon a perturbation, $\Delta\varepsilon$, of the host permittivity. Complex frequency shifts were calculated from direct solution of the TO dispersion relation ($\circ$ markers) and our GWS residue formula ($\times$ markers). Marker colour encodes $\Delta\varepsilon$ on a logarithmic scale. \textbf{(d)} Log–log plot of the magnitude of the frequency shift $|\Delta\omega|$ versus $\Delta\varepsilon$ for each EP.  Gradients of $1/2$, $1/3$, and $1/4$ are found for EP$_2$, EP$_3$ and EP$_4$ respectively (black dashed fits).}
	\label{fig:resonance_shifts}
\end{figure}

\emph{EPs in plasmonic nanowires--}As a final test of our theory, we investigate a class of plasmonic structures designed using transformation optics (TO) to support EPs of orders two, three, and four (denoted EP$_2$, EP$_3$, and EP$_4$ respectively) \cite{wang2025topological}. The corresponding compound nanowire and  dimer structures are depicted schematically in the insets of Figure~\ref{fig:resonance_shifts}(a)-(c). We introduce a perturbation in the background dielectric permittivity, $\epsilon_h\rightarrow\epsilon_h+\Delta\epsilon(\mathbf{r})$, and again track the resulting shifts in the complex resonance frequencies. Resonance shifts are computed first by directly solving the perturbed dispersion relation of the TO-based system \cite{wang2025topological} and secondly by applying \eqref{eq:domegaFinal} with the system's transfer matrix, whose poles also capture the EP, used in place of $\mathbf{S}$. Residue based predictions and numerically extracted root trajectories are shown in Figure~\ref{fig:resonance_shifts}(a)–(c) for each EP order. Good agreement between the GWS residue formula (crosses) and the TO-based solutions (circles) over six orders of magnitude variation in $\Delta\varepsilon$ (as depicted by marker color) is evident. Each EP of order $N$ also exhibits a $N$ fold symmetric splitting as predicted by \eqref{eq:domegaFinal}. Minor deviations arise at the largest perturbation amplitudes ($\Delta\varepsilon\sim10^{-4}$), particularly for higher-order EPs, which likely arise due to neglecting higher order terms in the perturbative expansion as well as potentially inaccuracies from the TO based approach, given the nanowire dimensions used. Plots of $|\Delta\omega|$ are also shown in Figure~\ref{fig:resonance_shifts}(d) confirming the theoretical $\Delta\varepsilon^{1/N}$ scaling (note that identical plots are found for each branch). 

In conclusion, we have developed a perturbative framework that extends prior  GWS perturbation theory to non-Hermitian systems exhibiting EPs and DPs. Our approach shares the merits of previous iterations and enables determination of internal resonant behavior from external scattering measurements in a flexible and straightforward way  \cite{Byrnes2024a,Byrnes2024b}. Our theory was validated through analytic modelling and electromagnetic simulations of assorted nanophotonic and TO engineered plasmonic systems. The final example in particular also demonstrates that although Eq.~(\ref{eq:domegaFinal}) was derived based on Eq.~(\ref{eq:mahauxS}), it does not depend on this exact relation and can also be used in situations where neither $\mathbf{H}$ nor $\mathbf{S}$ are convenient to describe. Our results demonstrate the method's accuracy, generality, and computational practicality, offering new perspectives and tools for sensitivity analysis, quantitative sensing, resonance control and inverse design of complex non-Hermitian environments (see e.g. \cite{Byrnes2024a,Byrnes2024b}).
	
\acknowledgements
K.W. and N.B were funded by a Nanyang Technological University Interdisciplinary Graduate Program (NTU-IGP) Scholarship and Singapore Ministry of Education Academic Research Fund (Tier 1) Grant RG66/23 respectively. M.R.F. was supported by IDMxS under the Singapore Ministry of Education Research Centres of Excellence scheme (EDUN C-33-18-279-V12) and by NTU Grant SUG:022824-00001.

\end{document}